# High-rate, Multi-Symbol-Decodable STBCs from Clifford Algebras

*Sanjay Karmakar*

Beceem Communications Pvt. Ltd.,
Bangalore.
skarmakar@beceem.com

*B. Sundar Rajan*

Department of ECE
Indian Institute of Science, Bangalore
bsrajan@ece.iisc.ernet.in

**Abstract**

It is well known that Space-Time Block Codes (STBCs) obtained from Orthogonal Designs (ODs) are single-symbol-decodable (SSD) and from Quasi-Orthogonal Designs (QODs) are double-symbol decodable. However, there are SSD codes that are not obtainable from ODs and DSD codes that are not obtainable from QODs. In this paper a method of constructing $g$-symbol decodable ($g$-SD) STBCs using representations of Clifford algebras are presented which when specialized to $g = 1, 2$ gives SSD and DSD codes respectively. For the number of transmit antennas $2^a$ the rate (in complex symbols per channel use) of the $g$-SD codes presented in this paper is $\frac{a+1-g}{2^{a-g}}$. The maximum rate of the DSD STBCs from QODs reported in the literature is $\frac{a}{2^{a-1}}$ which is smaller than the rate $\frac{a-1}{2^{a-2}}$ of the DSD codes of this paper, for $2^a$ transmit antennas. In particular, the reported DSD codes for 8 and 16 transmit antennas offer rates 1 and 3/4 respectively whereas the known STBCs from QODs offer only 3/4 and 1/2 respectively. The construction of this paper is applicable for any number of transmit antennas.

## 1. Introduction and Preliminaries

We consider a multiple antenna transmission system with $N_t$ number of transmit antennas and $N_r$ number of receive antennas. At each time slot $t$, the complex signals, $s_{ti}$, $i = 0, 1 \cdots, N_t - 1$ are transmitted from the $N_t$ transmit antennas simultaneously. Let $h_{ij} = \alpha_{ij} e^{\mathbf{j}\theta_{ij}}$ denote the path gain from the transmit antenna $i$ to the receive antenna $j$, where $\mathbf{j} = \sqrt{-1}$. Assuming that the path gain are constant over a frame length $N_t$ (we consider only square designs), the received signal $y_{tj}$ at the receive antenna $j$ at time $t$ is given by,

$$y_{tj} = \sum_{i=0}^{N_t-1} s_{ti} h_{ij} + n_{tj}, \quad (1)$$

for $j = 0, \cdots, N_r - 1$, $t = 0, \cdots, N_t - 1$, which in matrix notation is,

$$\mathbf{Y} = \mathbf{SH} + \mathbf{N} \quad (2)$$

where $\mathbf{Y} \in \mathbb{C}^{N_t \times N_r}$ is the received signal matrix, $\mathbf{S} \in \mathbb{C}^{N_t \times N_t}$ is the transmission matrix(also referred as codeword matrix), $\mathbf{N} \in \mathbb{C}^{N_t \times N_r}$ is the additive noise matrix and $\mathbf{H} \in \mathbb{C}^{N_t \times N_r}$ is the channel matrix, where $\mathbb{C}$ denotes the complex field. The entries of $\mathbf{H}$ are complex Gaussian with zero mean and unit variance and the entries of $\mathbf{N}$ are complex Gaussian with zero mean and variance $\sigma^2$. Both are assumed to be temporally and spatially white. We further assume that transmission power constraint is given by $E\left[tr\{\mathbf{SS}^H\}\right] = N_t^2$.

An $n \times n$ linear dispersion STBC [1] with $gK$ complex variables $x_1, x_2, \cdots, x_{gK}$, where $g$ and $K$ are positive integers, can be written as

$$\mathbf{S} = \sum_{i=1}^{K} S_i \quad (3)$$

where,

$$\begin{aligned} S_i &= \sum_{j=1}^{g} x_{(g(i-1)+j),I} A_{(g(i-1)+j),I} + \\ &+ x_{(g(i-1)+j),Q} A_{(g(i-1)+j),Q} \end{aligned} \quad (4)$$

and $x_l = x_{l,I} + jx_{l,Q} \in \mathcal{A}_l \subset \mathbb{C}$ for $1 \leq l \leq gK$ where $\mathcal{A}_l$ is the signal constellation from which the variable $x_i$ takes values. The set of $gK$ number of complex $n \times n$ matrices $A_j$, $1 \leq j \leq gK$ are called the weight matrices of the code and this set defines the code $\mathbf{S}$. With $|\mathcal{A}_i|$ denoting the number of points in the constellation, the rate of this code in bits per channel use is $R = \frac{1}{n}\sum_{i=1}^{gK} log_2(|\mathcal{A}_i|)$. Now assuming that perfect channel state information(CSI) is available at the receiver, the maximum likelihood (ML) decision rule minimizes the metric,

$$\mathbf{M(S)} \triangleq \min_{\mathbf{S}} tr((\mathbf{Y} - \mathbf{SH})^{\mathcal{H}}(\mathbf{Y} - \mathbf{SH})) = \parallel \mathbf{Y} - \mathbf{SH} \parallel^2. \quad (5)$$

It is clear that there are $\prod_{i=1}^{gK} |\mathcal{A}_i|$ different codewords and, in general, the ML decoding requires $\prod_{i=1}^{gK} |\mathcal{A}_i|$ computations, one for each codeword. But if the set of weight matrices are chosen such that the decoding metric (5) could be decomposed into,

$$\mathbf{M(S)} = \sum_{i=1}^{K} f_i(x_{(i-1)g+1}, x_{(i-1)g+2}, \cdots, x_{(i-1)g+g})$$

a sum of $K$ positive terms, each involving exactly $g$ complex variables only, then the decoding requires

$\sum_{i=1}^{K}\{\prod_{j=1}^{g}|\mathcal{A}_{j+(i-1)g}|\}$ computations and the code is called a *g-symbol decodable code* (*g*-SD code). The case $g = 1$ corresponds to Single-Symbol Decodable (SSD) codes that includes the well known codes from Orthogonal Designs (ODs) as a proper subclass, and have been extensively studied [2]-[13], few most recent ones being [12]-[13]. The codes corresponding to $g = 2$, are called *Double-Symbol-Decodable (DSD)* codes. The Quasi-Orthogonal Designs studied in [14]-[17] and [6] are proper subclass of DSD codes. If the weight matrices $A_j$, $1 \leq j \leq 2K$ of a *g*-SD code are all unitary then it is said to be a Unitary Weight DSD (UW-*g*-SD) code. The DSD codes of [14]-[17] are UW-DSD codes and that of [6] is not. Throughout this paper, we consider only UW codes.

The contributions of this paper are

- We obtain a set of sufficient conditions for a general linear dispersion STBC to be *g*-SD in terms of their weight matrices. Also, another set of sufficient conditions for the code to be *g*-SD is given which enables us to construct UW-*g*-SD codes from representations of Clifford algebras.

- For the number of transmit antennas $2^a$ the maximum rate (in complex symbols per channel use) of all the DSD codes reported in the literature with unitary weight matrices is $\frac{a}{2^{a-1}}$. Whereas we present UW-DSD codes with rate $\frac{a-1}{2^{a-2}}$ for $2^a$ transmit antennas. In particular, our code for 8 and 16 transmit antennas offer rates 1 and 3/4 respectively, whereas the known QODs offer only 3/4 and 1/2 respectively. The rate of our *g*-SD codes is $\frac{a+1-g}{2^{a-g}}$.

## 2. Sufficient conditions for Double-Symbol-Decodability

We begin with presenting a sufficient condition on the set of weight matrices of the code to be *g*-SD.

**Lemma 1** *The linear dispersion STBC given by* (3) *is a g-SD code if,*

$$S_l^H S_j + S_j^H S_l = 0, \ \forall \ 1 \leq l \neq j \leq K. \quad (6)$$

*Proof:* We see from (3) that,

$$\mathbf{S}^H \mathbf{S} = \Big(\sum_{i=1}^{K} S_i^H\Big)\Big(\sum_{i=1}^{K} S_i\Big). \quad (7)$$

If the conditions of (6) are satisfied, then it is easy to verify that,

$$\mathbf{S}^H \mathbf{S} = \sum_{i=1}^{K} S_i^H S_i. \quad (8)$$

Using (8) in (5) we get,

$$\mathbf{M}(\mathbf{S}) = Tr\Big[\sum_{i=1}^{K}(\mathbf{Y} - S_i\mathbf{H})^H(\mathbf{Y} - S_i\mathbf{H}) - (K-1)\mathbf{Y}^H\mathbf{Y}\Big]$$

$$= \sum_{i=1}^{K} Tr\Big[(\mathbf{Y} - S_i\mathbf{H})^H(\mathbf{Y} - S_i\mathbf{H})\Big] + \mathbf{M_c} \quad (9)$$

where $\mathbf{M_c} = Tr\Big[-(K-1)\mathbf{Y}^H\mathbf{Y}\Big]$. Note that in (9) the $\mathbf{M_c}$ term is same for all the codewords in the code book. Hence it is sufficient to minimize the first term only. But the first term,

$$\widetilde{\mathbf{M}}(\mathbf{S}) = \sum_{i=1}^{K} Tr\Big[(\mathbf{Y} - S_i\mathbf{H})^H(\mathbf{Y} - S_i\mathbf{H})\Big]$$

is a sum of $K$ square terms, each involving only $g$ complex variables. Hence the problem of ML decoding reduces to the problem of minimizing,

$$Tr\Big[(\mathbf{Y} - S_i\mathbf{H})^H(\mathbf{Y} - S_i\mathbf{H})\Big] \ \forall \ 1 \leq i \leq K. \quad (10)$$

Hence the code is *g*-SD if the conditions of (6) is satisfied. This completes the proof.

Next we derive a condition on weight matrices of the code so that (6) is satisfied. Towards this end, we denote,

$$\beta_i = \Big\{A_{g(i-1)+j,I}, A_{g(i-1)+j,Q}\Big\}_{j=1}^{g}, \ 1 \leq i \leq K. \quad (11)$$

A straight forward verification shows that

**Lemma 2** *Conditions of* (6) *is satisfied if the weight matrices of the code* (3) *satisfy the following condition,*

$$A^H B + B^H A = 0, \ \forall \ A \in \beta_i, B \in \beta_j, \text{ for } i \neq j. \quad (12)$$

We first introduce the notion of *normalizing a linear STBC* which not only simplifies the analysis of the codes but also provides deep insight into various aspects of different classes of codes discussed in this paper. Towards this end, let

$$S_U = \sum_{k=1}^{gK} x_{k,I} A'_{k,I} + x_{k,Q} A'_{k,Q} \quad (13)$$

be a UW-*g*-SD code. We normalize the weight matrices of the code as

$$\begin{aligned} A_{kI} &= A'^H_{1I} A'_{kI} \\ A_{kQ} &= A'^H_{1I} A'_{kQ}. \end{aligned} \ \forall \ 1 \leq k \leq gK \quad (14)$$

to get *the normalized version* of (13) to be

$$S_N = \sum_{k=1}^{gK} x_{k,I} A_{k,I} + x_{k,Q} A_{k,Q} \quad (15)$$

where $A_{1,I} = I_n$, the $n \times n$ identity matrix. We call the code $S_N$ to be the normalized code of $S_U$.

**Lemma 3** *The code $S_U$ is g-SD iff $S_N$ is g-SD. In other words normalization does not affect the DSD property.*

*Proof:* For $1 \le i_1 \ne i_2 \le K$, the equation of Lemma 2 is satisfied by the weight matrices of $S_U$ iff they are satisfied by the weight matrices of $S_N$ which is easily verified.

The following theorem identifies a set of sufficient conditions for a UW code to be UW-$g$-SD.

**Theorem 1** *A $n \times n$ UW code described by (13) and its normalized version given by (15) are UW-$g$-SD codes if the weight matrices of the normalized code satisfy the following conditions:*

$$A_{g(i-1)+1,I}^H = -A_{g(i-1)+1,I},\ 2 \le i \le K \quad (16)$$

$A_{g(i-1)+1,I}$ *and* $A_{g(j-1)+1,I}$ *anticommute*
$$\forall\, 2 \le i \ne j \le K \quad (17)$$

$A_{1,Q}, A_{2,I}, A_{2,Q}, \cdots, A_{g,I}, A_{g,Q}$ *are*

*Hermitian, commute among themselves and*

*with all the matrices* $A_{g(i-1)+1,I},\ 1 \le i \le K.$ (18)

*For all* $1 \le i \le K,$ *and* $2 \le j \le g$;

$$A_{g(i-1)+1,Q} = \pm A_{1,Q} A_{g(i-1)+1,I}, \quad (19)$$
$$A_{g(i-1)+1,I} = \pm A_{j,I} A_{g(i-1)+1,I}, \quad (20)$$
$$A_{g(i-1)+1,Q} = \pm A_{j,Q} A_{g(i-1)+1,I}, \quad (21)$$

*Proof:* Using the conditions (19), (20) and (21) of the theorem the sets (11) for $1 \le i_1 \ne i_2 \le K$ are

$$\beta_{i_1} = \left\{ \pm A_{j,I} A_{g(i_1)+1,I}, \pm A_{j,Q} A_{g(i_1)+1,Q} \right\}_{j=1}^{g}, \quad (22)$$

$$\beta_{i_2} = \left\{ \pm A_{j,I} A_{g(i_2)+1,I}, \pm A_{j,Q} A_{g(i_2)+1,Q} \right\}_{j=1}^{g}. \quad (23)$$

Let $x A_{g(i_1)+1,I} \in \beta_{i_1}$, $y A_{g(i_2)+1} \in \beta_{i_2}$ where $x, y \in \{I, \pm A_{1Q}, \pm A_{2I}, \pm A_{2Q}, \cdots \pm A_{(g-1),I}, \pm A_{(g-1),Q}\}$. Then

$$x A_{g(i_1)+1,I}{}^H y A_{gi_2+1,I} + y A_{g(i_2)+1,I}{}^H x A_{g(i_1)+1,I}$$
$$= A_{gi_1+1,I}{}^H x^H y A_{gi_2+1,I} + A_{gi_2+1,I}{}^H y^H x A_{gi_1+1,I}$$
$$= A_{gi_1+1,I}{}^H xy A_{gi_2+1,I} + A_{gi_2+1,I}{}^H yx A_{gi_1+1,I}$$
$$= A_{gi_1+1,I}{}^H xy A_{gi_2+1,I} + A_{gi_2+1,I}{}^H xy A_{gi_1+1,I}$$
$$= x A_{gi_1+1,I}{}^H A_{gi_2+1,I} y + x A_{gi_2+1,I}{}^H A_{gi_1+1,I} y$$
$$= x \left[ A_{gi_1+1,I}{}^H A_{gi_2+1,I} + A_{gi_2+1,I}{}^H A_{gi_1+1,I} \right] y$$
$$= x \left[ 0 \right] y = 0.$$

This completes the proof.

The requirements of Theorem 1 for UW-$g$SD can be easily stated using the Table 2 shown at the top of the next page as follows:

- The matrices of the first row should form a Hurwitz-Radon family of matrices
- The matrices of the first column should be
  - Hermitian
  - mutually commuting and
  - commute with all the matrices of the first row

**Definition 1** *A UW-DSD code satisfying the conditions of (16) is defined to be a Clifford Unitary Weight DSD (CUW-DSD) codes.*

The name in the above definition is due to the fact that such codes are constructable using matrix representations of real Clifford algebras as shown in the following section.

## 3. Construction of CUW-$g$-SD codes

In this section we present the construction of a new class of $2^a \times 2^a$ $g$-SD codes, the Clifford UW-$g$-SD (CUW-$g$-SD) codes, for $2^a$ transmit antennas, from unitary matrix representations of real Clifford algebras. For an excellent introduction to and basic properties of these representations see [4].

**Definition 2** *The Clifford algebra, denoted by $CA_L$, is the algebra over the real field $\mathbb{R}$ generated by $L$ objects $\gamma_k, \ k = 1, 2, \cdots, L$ which are anti-commuting, ($\gamma_k \gamma_j = -\gamma_j \gamma_k,\ \forall k \ne j,$) and squaring to $-1$, ($\gamma_k^2 = -1\ \forall k = 1, 2, \cdots, L$).*

A matrix representation of an algebra is completely specified by a representation of its generators. For a Clifford algebra, we are thus interested in unitary matrix representation of the generators $\gamma_k$'s. Let

$$\sigma_1 = \begin{bmatrix} 0 & 1 \\ -1 & 0 \end{bmatrix}, \sigma_2 = \begin{bmatrix} 0 & j \\ j & 0 \end{bmatrix} \text{ and } \sigma_3 = \begin{bmatrix} 1 & 0 \\ 0 & -1 \end{bmatrix} \quad (24)$$

and $\quad A^{\otimes m} = \underbrace{A \otimes A \otimes A \cdots \otimes A}_{m\ times}.$

From [4] we know that the representation of the generators of $CA_{2a+1}$ is given by

$$\begin{aligned}
R(\gamma_1) &= \pm j \sigma_3^{\otimes a} \\
R(\gamma_2) &= I_2^{\otimes a-1} \otimes \sigma_1 \\
R(\gamma_3) &= I_2^{\otimes a-1} \otimes \sigma_2 \\
&\ \vdots \\
R(\gamma_{2k}) &= I_2^{\otimes a-k} \otimes \sigma_1 \otimes \sigma_3^{\otimes k-1} \\
R(\gamma_{2k+1}) &= I_2^{\otimes a-k} \otimes \sigma_2 \otimes \sigma_3^{\otimes k-1} \\
&\ \vdots \\
R(\gamma_{2a}) &= \sigma_1 \otimes \sigma_3^{\otimes a-1} \\
R(\gamma_{2a+1}) &= \sigma_2 \otimes \sigma_3^{\otimes a-1}.
\end{aligned} \quad (25)$$

We add to this list the $2^a \times 2^a$ identity matrix, denoted by $I_{2^a}$, and designate it as $R(\gamma_0) = I_{2^a}$.

| $I = A_{1I}$ | $A_{g+1,I}$ | $A_{2g+1,I}$ | . | . | . | $A_{ig+1,I}$ | . | . | . | $A_{(K-1)g+1,I}$ |
|---|---|---|---|---|---|---|---|---|---|---|
| $A_{1Q}$ | $A_{1Q}A_{g+1,I}$ | $A_{1Q}A_{2g+1,I}$ | . | . | . | $A_{1Q}A_{ig+1,I}$ | . | . | . | $A_{1Q}A_{(K-1)g+1,I}$ |
| $A_{2I}$ | $A_{2I}A_{g+1,I}$ | $A_{2I}A_{2g+1,I}$ | . | . | . | $A_{2I}A_{ig+1,I}$ | . | . | . | $A_{2I}A_{(K-1)g+1,I}$ |
| $A_{2Q}$ | $A_{2Q}A_{g+1,I}$ | $A_{2Q}A_{2g+1,I}$ | . | . | . | $A_{2Q}A_{ig+1,I}$ | . | . | . | $A_{2Q}A_{(K-1)g+1,I}$ |
| . | . | . | . | . | . | . | . | . | . | . |
| . | . | . | . | . | . | . | . | . | . | . |
| . | . | . | . | . | . | . | . | . | . | . |
| $A_{gI}$ | $A_{gI}A_{g+1,I}$ | $A_{gI}A_{2g+1,I}$ | . | . | . | $A_{gI}A_{ig+1,I}$ | . | . | . | $A_{gI}A_{(K-1)g+1,I}$ |
| $A_{gQ}$ | $A_{gQ}A_{g+1,I}$ | $A_{gQ}A_{2g+1,I}$ | . | . | . | $A_{gQ}A_{ig+1,I}$ | . | . | . | $A_{gQ}A_{(K-1)g+1,I}$ |

### 3.1. Construction of CUW-$g$-SD codes

To construct CUW-$g$-SD codes from the last $2g$ matrices $\{R(\gamma_{2(a+1-g)}), R(\gamma_{2(a+1-g)+1}), \cdots, R(\gamma_{2a}), R(\gamma_{2a+1})\}$ we construct the following $2g-1$ new matrices,

$$\begin{aligned}
\alpha_1 &= \pm jR(\gamma_{2a-2})R(\gamma_{2a-1}) \\
\alpha_2 &= \pm jR(\gamma_{2a})R(\gamma_{2a+1}) \\
&\vdots \quad \vdots \\
\alpha_g &= \pm jR(\gamma_0)R(\gamma_1) \\
\alpha_{g+1} &= \pm \alpha_1 \alpha_2 \\
\alpha_{g+2} &= \pm \alpha_3 \alpha_4 \\
&\vdots \quad \vdots \\
\alpha_{g+\frac{g}{2}} &= \pm \alpha_{g-1} \alpha_g \\
\alpha_{g+\frac{g}{2}+1} &= \pm \alpha_{g+1} \alpha_{g+2} \\
\alpha_{g+\frac{g}{2}+2} &= \pm \alpha_{g+3} \alpha_{g+4} \\
&\vdots \quad \vdots \\
\alpha_{g+\frac{g}{2}+\frac{g}{4}} &= \pm \alpha_{g+\frac{g}{2}-1} \alpha_{g+\frac{g}{2}} \\
&\vdots \quad \vdots \\
\alpha_{2g-1} &= \Pi_{i=1}^{g} \alpha_g.
\end{aligned} \quad (26)$$

Note that the set of matrices $\{\alpha_i\}_{i=1}^{2g-1}$ have the following properties: (i) They are mutually commuting, (ii) Hermitian and (iii) each commutes with $R(\gamma_1), R(\gamma_2), \cdots, R(\gamma_{2a-2g+1})$. In the above construction of $\alpha_i$s there is nothing special about the last $2g$ matrices. Any $2g$ from the set $\{R(\gamma_1), R(\gamma_2), \cdots, R(\gamma_{2a+1})\}$ could have been selected. Next, for $1 \leq i \leq 2a-2g+2$, we construct the weight matrices of the code combining the matrices defined above in the following way,

$$\left.\begin{aligned}
A_{g(i-1)+1,I} &= R(\gamma_{i-1}) \\
A_{g(i-1)+2,I} &= \pm \alpha_1 R(\gamma_{i-1}) \\
&\vdots \quad \vdots \\
A_{g(i-1)+g,I} &= \pm \alpha_{g-1} R(\gamma_{i-1}) \\
A_{g(i-1)+1,Q} &= \pm \alpha_g R(\gamma_{i-1}) \\
A_{g(i-1)+2,Q} &= \pm \alpha_{g+1} R(\gamma_{i-1}) \\
&\vdots \quad \vdots \\
A_{g(i-1)+g,Q} &= \pm \alpha_{2g-1} R(\gamma_{i-1})
\end{aligned}\right\} \quad (27)$$

**Theorem 2** *The $2^a \times 2^a$ code with the weight matrices given by* (27) *is a CUW-$g$-SD code.*

*Proof* It is easily checked that the weight matrices satisfy the conditions of Theorem 1 and hence the code is CUW-$g$-SD.

**Corollary 1** *The rate of the $2^a \times 2^a$ CUW-$g$-SD codes of Theorem 2 is $\frac{a+1-g}{2^{a-g}}$ complex symbols per channel use.*

**Example 1** *Let the representation matrices of the $CA_{2a+1}$, where $a = 3$ be,*

$$\begin{aligned}
R(\gamma_0) &= I_2 \otimes I_2 \times I_2, R(\gamma_1) = I_2 \otimes I_2 \times \sigma_1, \\
R(\gamma_2) &= I_2 \otimes I_2 \times \sigma_2, R(\gamma_3) = I_2 \otimes \sigma_1 \times \sigma_3, \\
R(\gamma_4) &= I_2 \otimes \sigma_2 \times \sigma_3, R(\gamma_5) = \sigma_1 \otimes \sigma_3 \times \sigma_3 \\
R(\gamma_6) &= \sigma_2 \otimes \sigma_3 \times \sigma_3, R(\gamma_7) = j\sigma_3 \otimes \sigma_3 \times \sigma_3
\end{aligned}$$

*Now according to the prescription of the construction procedure,*

$$\begin{aligned}
\alpha_1 &= jR(\gamma_4)R(\gamma_5) = \sigma_1 \otimes \sigma_1 \times I_2 \\
\alpha_2 &= jR(\gamma_6)R(\gamma_7) = j\sigma_1 \otimes I_2 \times I_2 \\
\alpha_3 &= R(\gamma_4)R(\gamma_5)R(\gamma_6)R(\gamma_7) = jI_2 \otimes \sigma_1 \times I_2
\end{aligned}$$

*and the weight matrices given by* (28) *at the top of the next page. Now if we construct the code, we get the code given by* (29) *at the top of the next page.*

Note that for 8 transmit antennas our CUW-DSD code achieves rate 1 whereas all known QODs with unitary weight matrices for 8 transmit antennas achieve rate only $\frac{3}{4}$. However, the 8 transmit antenna DSD code with non-unitary weight matrices of [6] achieve rate 1, but has larger Peak-to-Average Power Ratio due to the presence of zero entries in the code, compared to our CUW-DSD codes.

### 4. Acknowledgements

The authors would like to thank G.Susinder Rajan for useful discussions on the subject of this paper. This work was partly supported by the DRDO-IISc Program on Advanced Research in Mathematical Engineering and by the Council of Scientific & Industrial Research (CSIR), India, through Research Grant (22(0365)/04/EMR-II) to B.S. Rajan.

$$\begin{aligned}
A_{1I} &= R(\gamma_0) = I_2 \otimes I_2 \times I_2 & A_{3I} &= R(\gamma_1) = I_2 \otimes I_2 \times \sigma_1 & A_{5I} &= R(\gamma_2) = I_2 \otimes I_2 \times \sigma_2 & A_{7I} &= R(\gamma_3) = I_2 \otimes \sigma_1 \times \sigma_3 \\
A_{1Q} &= \alpha_1 A_{1I} = \sigma_1 \otimes \sigma_1 \times I_2 & A_{3Q} &= \alpha_1 A_{3I} = \sigma_1 \otimes \sigma_1 \times \sigma_1 & A_{5Q} &= \alpha_1 A_{5I} = \sigma_1 \otimes \sigma_1 \times \sigma_2 & A_{7Q} &= \alpha_1 A_{7I} = -\sigma_1 \otimes I_2 \times \sigma_3 \\
A_{2I} &= \alpha_2 R(\gamma_0) = j\sigma_1 \otimes I_2 \times I_2 & A_{4I} &= \alpha_2 R(\gamma_1) = j\sigma_1 \otimes I_2 \times \sigma_1 & A_{6I} &= \alpha_2 R(\gamma_2) = j\sigma_1 \otimes I_2 \times \sigma_2 & A_{8I} &= \alpha_2 R(\gamma_3) = j\sigma_1 \otimes \sigma_1 \times \sigma_3 \\
A_{2Q} &= \alpha_3 R(\gamma_0) = jI_2 \otimes \sigma_1 \times I_2 & A_{4Q} &= \alpha_3 R(\gamma_1) = jI_2 \otimes \sigma_1 \times \sigma_1 & A_{6Q} &= \alpha_3 R(\gamma_2) = jI_2 \otimes \sigma_1 \times \sigma_2 & A_{8Q} &= \alpha_3 R(\gamma_3) = -jI_2 \otimes I_2 \times \sigma_3
\end{aligned} \quad (28)$$

$$\begin{bmatrix}
x_{1I} - jx_{8Q} & x_{3I} + jx_{5I} & x_{7I} + jx_{2Q} & -x_{6Q} + jx_{4Q} & -x_{7Q} + jx_{2I} & -x_{6I} + jx_{4I} & x_{1Q} + jx_{8I} & x_{3Q} + jx_{5Q} \\
-x_{3I} + jx_{5I} & x_{1I} + jx_{8Q} & -x_{6Q} - jx_{4Q} & -x_{7I} + jx_{2Q} & -x_{6I} - jx_{4I} & x_{7Q} + jx_{2I} & -x_{3Q} + jx_{5Q} & x_{1Q} - jx_{8I} \\
-x_{7I} - jx_{2Q} & x_{6Q} - jx_{4Q} & x_{1I} - jx_{8Q} & x_{3I} + jx_{5I} & -x_{1Q} - jx_{8I} & -x_{3Q} - jx_{5Q} & -x_{7Q} + jx_{2I} & -x_{6I} + jx_{4I} \\
x_{6Q} + jx_{4Q} & x_{7I} - jx_{2Q} & -x_{3I} + jx_{5I} & x_{1I} + jx_{8Q} & x_{3Q} - jx_{5Q} & -x_{1Q} + jx_{8I} & -x_{6I} - jx_{4I} & x_{7Q} + jx_{2I} \\
x_{7Q} - jx_{2I} & x_{6I} - jx_{4I} & -x_{1Q} - jx_{8I} & -x_{3Q} - jx_{5Q} & x_{1I} - jx_{8Q} & x_{3I} + jx_{5I} & x_{7I} + jx_{2Q} & -x_{6Q} + jx_{4Q} \\
x_{6I} + jx_{4I} & -x_{7Q} - jx_{2I} & x_{3Q} - jx_{5Q} & -x_{1Q} + jx_{8I} & -x_{3I} + jx_{5I} & x_{1I} + jx_{8Q} & -x_{6Q} - jx_{4Q} & -x_{7I} + jx_{2Q} \\
x_{1Q} + jx_{8I} & x_{3Q} + jx_{5Q} & x_{7Q} - jx_{2I} & x_{6I} - jx_{4I} & -x_{7I} - jx_{2Q} & x_{6Q} - jx_{4Q} & x_{1I} - jx_{8Q} & x_{3I} + jx_{5I} \\
-x_{3Q} + jx_{5Q} & x_{1Q} - jx_{8I} & x_{6I} + jx_{4I} & -x_{7Q} - jx_{2I} & x_{6Q} + jx_{4Q} & -x_{7I} - jx_{2Q} & -x_{3I} + jx_{5I} & x_{1IQ} + jx_{8Q}
\end{bmatrix} \quad (29)$$

## 5. References

[1] B. Hassibi and B. Hochwald, "High-rate codes that are linear in space and time," *IEEE Trans. Inform. Theory*, vol.48, no.7, pp.1804-1824, July 2002.

[2] S.M. Alamouti, "A simple transmitter diversity scheme for wireless communications," *IEEE J. Select Areas Comm.,* vol 16 pp. 1451-1458, Oct. 1998.

[3] V. Tarokh, H. Jafarkhani and A. R. Calderbank, "Space-Time block codes from orthogonal designs," *IEEE Trans. Inform. Theory,* vol.45, pp.1456-1467, July 1999.

[4] O. Tirkkonen and A. Hottinen, "Square-matrix embeddable space-time block codes for complex signal constellations," *IEEE Trans. Inform. Theory*, vol.48, no.2, Feb. 2002, pp.384-395.

[5] G. Ganesan and P. Stoica, "Space-time diversity using orthogonal and amicable orthogonal designs," in *Proc. IEEE Int. Conf. Acoustics, Speech and Signal Processing* (ICASSP 2000), Istanbul, Turkey, 2000, pp. 2561-2564.

[6] Md. Zafar Ali Khan, B. Sundar Rajan and M.H.Lee,"On Single-Symbol and Double-Symbol decodable STBCs," Proc. ISIT 2003, Yokohama, Japan, June 29-July 4, 2003, p.127.

[7] Md. Zafar Ali Khan and B. Sundar Rajan,"Single-Symbol Maximum-Likelihood Decodable Linear STBCs," *IEEE Transactions on Information Theory,* vol. 52, no.5, May 2006, pp.2062-2091.

[8] H.Wang, D.Wang and X-G.Xia,"On optimal QOSTBC with minimal decoding complexity," Submitted to IEEE transactions on Information Theory.

[9] H.Wang, D.Wang and X-G.Xia,"On Optimal Quasi-orthogonal space-time block codes with minimum decoding complexity," Proc. ISIT 2005, Adelaide, Nov. 2005, pp.1168-1172.

[11] C.Yuen, Y.L. Guan and T.T. Tjhung, " Full rate full diversity STBC with constellation rotation," *Proc. VTC 2003,* Spring vol. 1, pp. 296-300, Seogwipo, Korea, April 2003.

[10] G.Yuen, Y.L.Guan and T.T.Tjhung, "Construction of quasi-orthogonal STBC with minimum decoding complexity," Proc. ISIT 2004, Chicago, June/July 2004, p.308.

[11] G.Yuen, Y.L.Guan and T.T.Tjhung, "Optimizing quasi-orthogonal STBC with group-constrained linear transformation," Proc. GLOBECOM 2004, Dallas, Texas, Dec. 2004, pp.550-554.

[12] Sanjay Karmakar and B. Sundar Rajan, "Minimum-decoding-complexity maximum-rate space-time block codes from Clifford algebras," Proc. ISIT 2006, Seattle, USA, July 09-14, 2006, pp.788-792.

[13] Sanjay Karmakar and B. Sundar Rajan, " Non-unitary weight space-time block codes with minimum decoding complexity," Proc. ISIT 2006, Seattle, USA, July 09-14, 2006, pp.793-797.

[14] Jafarkhani, H., "A quasi-orthogonal space-time block code," IEEE Transactions on Communications, Volume 49, Issue 1, Jan. 2001, pp.1-4.

[15] Weifeng Su and X.G.Xia, "Signal constellations for QOSTBC with full diversity," IEEE Trans. Inform. Theory, Vol-50, Oct, 2004, pp.2331-2347.

[16] O.Tirkkonen, A.Boariu and A.Hottinen, "Minimal and non-orthogonality rate 1 space time block code for 3+ Tx antennas," IEEE 6th Int. Symp. on Spread Spectrum Tech. and Appl. (ISSSTA 2000), pp.429-432.

[17] N.Sharma and C.B.Papadias, "Improved Quasi-orthogonal codes," in Proc. IEEE Wireless Communications and Networking Conference (WCNC 2002), March 17-21, vol.1, pp.169-171.